\documentclass[twocolumn]{aastex61}

\newcommand{\rii}{\emph{r}-II}
\newcommand{\ri}{\emph{r}-I}
\newcommand{\rp}{\emph{r}-process}
\newcommand{\fullname}{2MASS~J09544277$+$5246414}
\newcommand{\shortname}{J0954$+$5246}
\newcommand{\teff}{T_{\mathrm{eff}}}
\newcommand{\feh}{[\mathrm{Fe/H}]}

\newenvironment{tightcenter}{%
  \setlength\topsep{0pt}
  \setlength\parskip{0pt}
  \begin{center}
}{%
  \end{center}
}

\usepackage{graphicx}
\usepackage{amsmath}
\usepackage{hyperref,enumerate}
\usepackage{natbib}

\hypersetup{
  unicode=false,         
  pdftoolbar=true,        
  pdfmenubar=true,        
  pdffitwindow=true,       
  pdfstartview={FitH},      
  pdftitle={J095442},   
  pdfauthor={emholmbeck},      
  pdfsubject={Astronomy},     
  pdfcreator={dvipdf},      
  pdfproducer={dvipdf},      
  pdfkeywords={metal-poor stars}, 
  pdfnewwindow=true,       
  colorlinks=true,        
  linkcolor=purple,         
  citecolor=blue,         
  filecolor=magenta,       
  urlcolor=cyan,         
  breaklinks=true,
  linktocpage
}

\shorttitle{\shortname: A Highly Actinide-Enhanced \emph{\textsc{r}}-II Star in the Milky Way}
\shortauthors{Holmbeck et al.}

\begin{document}

\title{The \emph{R}-Process Alliance: \fullname, The Most \\
Actinide-Enhanced \emph{r}-II Star Known}

\author{Erika M.\ Holmbeck}
\affiliation{Department of Physics, University of Notre Dame, Notre Dame, IN 46556, USA}
\affiliation{JINA Center for the Evolution of the Elements, USA}

\author{Timothy C.\ Beers}
\affiliation{Department of Physics, University of Notre Dame, Notre Dame, IN 46556, USA}
\affiliation{JINA Center for the Evolution of the Elements, USA}

\author{Ian U.\ Roederer}
\affiliation{Department of Astronomy, University of Michigan, 1085 S. University Ave., Ann Arbor, MI 48109, USA}
\affiliation{JINA Center for the Evolution of the Elements, USA}

\author{Vinicius M.\ Placco}
\affiliation{Department of Physics, University of Notre Dame, Notre Dame, IN 46556, USA}
\affiliation{JINA Center for the Evolution of the Elements, USA}

\author{Terese T.\ Hansen}
\affiliation{The Observatories of the Carnegie Institution of Washington, Pasadena, CA 91101, USA}

\author{Charli M.\ Sakari}
\affiliation{Department of Astronomy, University of Washington, Seattle, WA 98195, USA}

\author{Christopher Sneden}
\affiliation{Department of Astronomy and McDonald Observatory, The University of Texas, Austin, TX 78712, USA}

\author{Chao Liu}
\affiliation{Key Lab of Optical Astronomy, National Astronomical Observatories, CAS, 100012, Beijing, China}

\author{Young Sun Lee}
\affiliation{Department of Astronomy and Space Science, Chungnam National University, Daejeon 34134, Korea}

\author{John J.\ Cowan}
\affiliation{HLD Department of Physics \& Astronomy, University of Oklahoma, Norman, OK 73019, USA}

\author{Anna Frebel}
\affiliation{Department of Physics and Kavli Institute for Astrophysics and Space Research, Massachusetts Institute of Technology, Cambridge, MA 02139, USA}
\affiliation{JINA Center for the Evolution of the Elements, USA}

\correspondingauthor{Erika M.\ Holmbeck}
\email{eholmbec@nd.edu}

\begin{abstract}

We report the discovery of a new actinide-boost star, \fullname, originally identified as a very bright ($V = 10.1$), extremely metal-poor ($\feh=-2.99$) K giant in the LAMOST survey, and found to be highly \rp-enhanced (\rii; [Eu/Fe]$=+1.28$]), during the snapshot phase of the \emph{R}-Process Alliance (RPA).
Based on a high S/N, high-resolution spectrum obtained with the Harlan J. Smith 2.7-m telescope, this star is the first confirmed actinide-boost star found by RPA efforts.
With an enhancement of $\textrm{[Th/Eu]}=+0.37$, \fullname\ is also the most actinide-enhanced \rii\ star yet discovered, and only the sixth metal-poor star with a measured uranium abundance ($\textrm{[U/Fe]}=+1.40$).
Using the Th/U chronometer, we estimate an age of 13.0$\pm$4.7~Gyr for this star.
The unambiguous actinide-boost signature of this extremely metal-poor star, combined with additional \rp-enhanced and actinide-boost stars identified by the RPA, will provide strong constraints on the nature and origin of the \rp\ at early times. 
\end{abstract}


\received{2018 April 23}
\revised{2018 May 14}
\accepted{2018 May 14}
\submitjournal{\apjl}

\section{Introduction}
First described by \citet{b2fh} and \citet{cameron1957}, the rapid neutron-capture (``\emph{r}-") process is the physical mechanism responsible for synthesizing roughly half of the elements heavier than iron present in the Solar System.
The astrophysical site of the \rp\ is currently a topic of curiosity and debate, but recent follow-up observations on the gravitational wave event GW170817 \citep{abbott2017} strongly support the merging of two neutron stars in a binary system as one possible production site. 
Photometric \citep{drout2017} and spectroscopic \citep{shappee2017} evidence for the presence of unstable lanthanide isotopes produced by the \rp\ have been identified from observations of the kilonova source SSS17a associated with the neutron star merger \citep{kilpatrick2017,cowperthwaite2017}. 
Events such as magneto-rotational instability (``jet") supernovae also remain a viable astrophysical site for the ``main" \rp\ \citep[see, e.g.,][]{nishimura2017}.

Individual \rp-enhanced stars in the halo of the Milky Way have recorded in their photospheres the nucleosynthetic products of \rp\ events that occurred early in Galactic history.
These stars are characterized by their typically low metallicities ($\feh \lesssim -2$) and moderate to large over-abundances of \rp\ elements.
We use the designations ``\ri": $+0.3\leq \rm{[Eu/Fe]}\leq +1.0$ and ``\rii": $\rm{[Eu/Fe]} > +1.0$ to quantify the observed level of \rp\ enhancement \citep{beers2005}.
The \rp\ elements present in the photospheres of \rp-enriched stars likely reflect the contributions from one to at most a few \rp\ events preceding their formation.

About 40 \rii\ stars are currently known in the Galactic halo; ten of which were recently identified through an ongoing search for \rp-enhanced stars, conducted as part of the \emph{R}-Process Alliance \citep[RPA;][]{sakari2018,hansen2018}.
Another seven \rii\ stars have been identified in the ultra faint dwarf (UFD) galaxy Reticulum II \citep{ji2016,roederer2016}.
The main \rp\ elemental-abundance pattern (Ba to Hf) of \ri\ and \rii\ stars is strikingly homogeneous and closely follows the \rp\ pattern observed in the Solar System for stars covering a wide range of metallicities \citep{sneden2002,siqueira2014,hansen2017}.
The universality of the main \rp\ pattern among Milky Way halo stars, UFDs, and classical dwarf galaxies suggests that
either the \rp\ behaves uniformly regardless of its astrophysical site, or that the observed \rp\ material comes from one type of \rp-element production event.

The identification and analysis of metal-poor, \rp-enhanced stars is essential for elucidating the nature and origin of the astrophysical \rp.
Although the elemental-abundance patterns of \rp-enriched stars are relatively homogeneous among the lanthanides, the actinide abundances vary significantly; about 30\% of \rp-enhanced stars exhibit Th/Eu abundance ratios a factor of 2--3 higher than the majority of other \rp-enriched stars \citep[see][]{hill2002,mashonkina2014}. 
The mechanism for producing this ``actinide boost" is neither well-studied nor well-understood and suffers from low number statistics.
Studying actinide-boost stars can provide clues to the different astrophysical conditions necessary to produce elements beyond the third \rp\ peak and help distinguish between suggested astrophysical sites for the \rp.

In this letter, we present the elemental-abundance analysis, based on high-resolution spectroscopic follow-up conducted as part of the RPA, of a newly-identified actinide-boost \rii\ star: \fullname\ (hereafter \shortname).
Not only is \shortname\ the sixth metal-poor star with a uranium measurement, it is also the brightest \citep[$V=10.095$; APASS;][]{henden2015} and the most actinide-enhanced ($\textrm{[Th/Eu]}=+0.37$) \rii\ star currently known.
Due to its brightness, extremely low metallicity, and lack of carbon enhancement, we are able to report measurements for 42 elements available from ground-based, optical observations.

\section{Observations and Analysis}

	\begin{table}[t]
    {\footnotesize
	\begin{tightcenter}
	\caption{Derived Parameters for \shortname.\label{tab:params}}
	\begin{tabular}{l r r r}\hline\hline
	Parameter  &  \multicolumn{1}{c}{LAMOST}  &  \multicolumn{1}{c}{SSPP}  &  \multicolumn{1}{c}{TS-23\tablenotemark{a}} \\ \hline
	$\teff$ (K)  &  4462 $\pm$ 110  &  4340 $\pm$ 150  &  4340 $\pm$ 125 \\
	$\log g$ (cgs)  &  0.91 $\pm$ 0.19  &  0.6 $\pm$ 0.30  &  0.41 $\pm$ 0.20 \\
	$\feh$  &  $-$2.46 $\pm$ 0.11  &  $-$3.16 $\pm$ 0.15  &  $-$2.99 $\pm$ 0.10 \\
	$\xi$ (km~s$^{-1}$)  &  ---  &  ---  &  2.28 $\pm$ 0.20 \\
	$[$C/Fe$]$ &  ---  &  $-$0.34  &  $-$0.50 $\pm$ 0.20 \\
	MJD  &  57030  &  57030  &  58128 \\
	RV (km~s$^{-1}$)  &  $-$71 $\pm$ 5  &  $-$71.9 $\pm$ 1.8  &  $-$67.7 $\pm$ 0.1 \\ 
	\hline
	\end{tabular}
	\end{tightcenter}
    \tablenotetext{a}{Final parameters adopted for this work.}
    }
	\end{table}

\shortname\ was first identified as a candidate very metal-poor K giant in the LAMOST (DR4) Survey \citep{liu2014}.
The SEGUE Stellar Parameter Pipeline \citep[SSPP;][]{lee2008a, lee2008b} was used to estimate the atmospheric parameters ($\teff$, $\log g$, and $\feh$)---as well as its carbon-to-iron ratio ([C/Fe], as described in \citealt{lee2013})---from the LAMOST medium-resolution data.
The metallicity and [C/Fe] estimated by the SSPP revealed \shortname\ to be extremely metal-poor, non-carbon-enhanced, and with a low effective temperature, suitable for inclusion in the RPA search for \rp-enhanced stars.

We carried out high-resolution (``portrait") spectroscopic observations during 2018A using the Harlan J. Smith 107-in (2.7-m) telescope and the TS23 echelle spectrograph \citep{tull1995} at McDonald Observatory.
The high-resolution setup uses a 1$\farcs$2 slit and 1$\times$1 binning, yielding a resolving power of $R\sim 60,000$, with full wavelength coverage of 3600--5800\,{\AA} and partial wavelength coverage up to 10,000\,{\AA}. 
From co-addition of nine spectra (total exposure time 15,600~s), a final S/N of 90 per resolution element at 4100\,{\AA} was achieved.
The data were reduced using standard IRAF packages \citep{tody1993}.
    
Equivalent widths (EWs) of 125 \ion{Fe}{1} and 27 \ion{Fe}{2} lines were measured using the \texttt{splot} task in IRAF, fitting a Gaussian profile to each line, and deblending where necessary.
Individual \ion{Fe}{1} and \ion{Fe}{2} abundances were derived from their EWs using the current version of the LTE stellar line analysis code \texttt{MOOG}\protect\footnote{\url{https://github.com/alexji/moog17scat}} \citep{sneden1973}, which includes an appropriate treatment of scattering \citep{sobeck2011}.
We use $\alpha$-enhanced ($[\alpha/\textrm{Fe}]=0.4$) ATLAS9 model atmospheres \citep{castelli2004}.
A list of all Fe lines used to determine the atmospheric parameters of \shortname\ is provided in a .tar.gz package.

All other elemental abundances were derived from spectral synthesis of features in the portrait spectrum using \texttt{MOOG}.
Oscillator strengths and excitation potentials for all lines in this work were generated with \texttt{linemake}\protect\footnote{\url{https://github.com/vmplacco/linemake}}, which compiles recent, accurate atomic transition data and includes hyperfine splitting.
Isotopic ratios, employed when synthesizing features with hyperfine-splitting and isotopic shift effects, were taken from the Solar \rp\ ratios in \citet{sneden2008}.

\shortname\ is a cool, extremely metal-poor giant; at such extremes, it becomes difficult to determine atmospheric parameters based on spectroscopy alone.
Over a large range of excitation potentials, the spectroscopically-derived \ion{Fe}{1} abundances equilibrate at 4100\,K.
However, \ion{Fe}{1} and \ion{Fe}{2} abundance equilibration fails at these low effective temperatures, disagreeing by about 0.3~dex.
Instead of a pure spectroscopic approach, we first use the temperature scaling for giant stars from \citet{alonso1999}, applied to 2MASS photometry \citep{skrutskie2006}, to obtain a $\teff$ estimate of 4340\,K, based on the $(J-K)$ color of $0.716$.\footnote{This photometric temperature matches that obtained from the SSPP, based on the medium-resolution spectrum.}
After applying the correction from \citet{frebel2013} to the 4100\,K found previously, we find a spectroscopic temperature of 4360\,K, which agrees with the photometric estimate and validates the spectroscopic method, despite the differences that arise between \ion{Fe}{1} and \ion{Fe}{2} abundances at low temperatures.
The reddening is sufficiently low \citep[$E(B-V)$ = 0.007;][]{schlafly2011} that it does not significantly affect the ($J-K$) color.
After fixing the temperature to the photometric estimate of 4340\,K, ionization equilibration was carried out to find the surface gravity and metallicity.
Microturbulence was determined by minimizing the \ion{Fe}{1} and \ion{Fe}{2} abundance trends in reduced equivalent width.
Carbon abundance was measured based on high-resolution spectral synthesis of the 4300~\AA\ CH $G$-band.
Table~\ref{tab:params} summarizes the derived atmospheric parameters (and their estimated errors) for this star from the medium- and high-resolution spectra.
Measured radial velocities are also reported in Table~\ref{tab:params}; based on the similarity between radial velocity measurements, there is no indication of binarity for \shortname.

\section{The \emph{r}-Process Pattern}

	\begin{figure*}[t]
	 \begin{center}
	 \centerline{\includegraphics[width=0.78\textwidth]{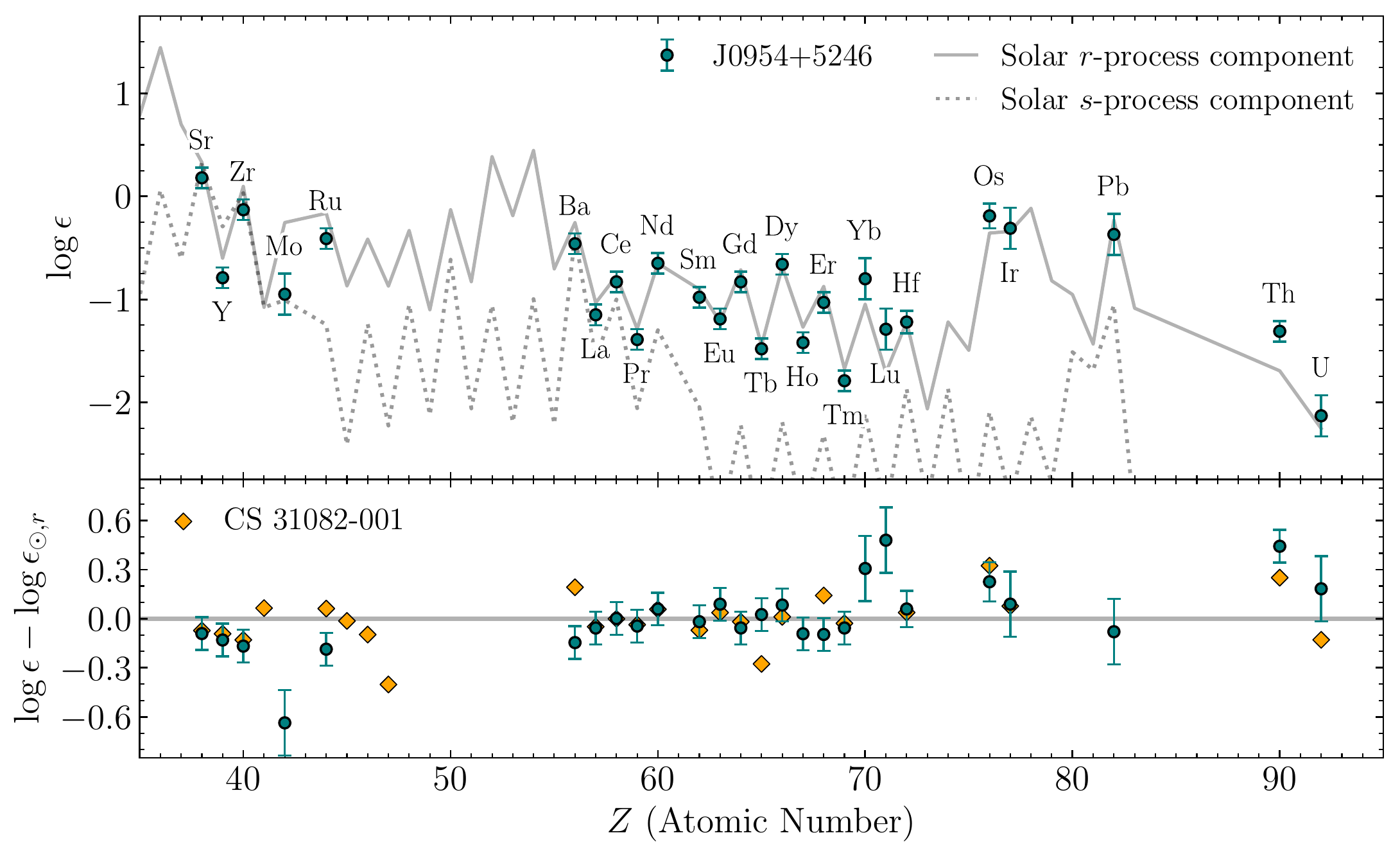}}
	 \caption{\label{fig:pattern} Top panel: Full \rp\ elemental-abundance pattern of \shortname. Also shown are the scaled-Solar \emph{s}-process abundances (dotted line). The Solar \rp\ component (solid line) is the abundance residual from the \emph{s}-process. Both components are normalized to Sr. Solar abundances are from \citet{asplund2009}; \emph{s}-process abundance fractions are from \citet{arlandini1999}. Bottom panel: Difference between the observed abundances and the Solar \rp\ component for \shortname\ and observations of the canonical actinide-boost star {CS~31082--001}, as reported by \citet{hill2002}. The residuals in the bottom panel are normalized to the average residual between Ba and Tm.}
	 \end{center}
	\end{figure*}

    \begin{table}
    {\scriptsize
    \caption{Derived elemental abundances of \shortname.\label{tab:abundances}}
    \begin{tightcenter}
    \begin{tabular}{lrrrrrr}
    \hline\hline
    Species & $\log\epsilon_{\odot} $ & $\log\epsilon$ & [X/H] & [X/Fe] & $\sigma_{\log \epsilon}$ & $N$ \\
    \hline
    C (CH)  &  8.43  &  4.94  &  $-$3.49  &  $-$0.50\tablenotemark{a}  &  0.20  &  1 \\ 
    \ion{Na}{1}  &  6.24  &  3.53  &  $-$2.72  &  0.28  &  0.10  &  2 \\ 
    \ion{Mg}{1}  &  7.60  &  5.25  &  $-$2.35  &  0.64  &  0.10  &  4 \\ 
    \ion{Al}{1}  &  6.45  &  3.06  &  $-$3.39  &  $-$0.40  &  0.20  &  1 \\ 
    \ion{Ca}{1}  &  6.34  &  3.60  &  $-$2.74  &  0.25  &  0.10  &  14 \\ 
    \ion{Sc}{2}  &  3.15  &  $-$0.02  &  $-$3.17  &  $-$0.18  &  0.10  &  7 \\ 
    \ion{Ti}{1}  &  4.95  &  1.97  &  $-$2.98  &  0.01  &  0.10  &  9 \\ 
    \ion{Ti}{2}  &  4.95  &  2.17  &  $-$2.78  &  0.21  &  0.10  &  11 \\ 
    \ion{V}{2}  &  3.93  &  0.94  &  $-$2.99  &  0.00  &  0.10  &  2 \\ 
    \ion{Cr}{1}  &  5.64  &  2.34  &  $-$3.30  &  $-$0.31  &  0.10  &  7 \\ 
    \ion{Cr}{2}  &  5.64  &  2.81  &  $-$2.84  &  0.16  &  0.10  &  2 \\ 
    \ion{Mn}{1}  &  5.43  &  1.64  &  $-$3.79  &  $-$0.80  &  0.14  &  3 \\ 
    \ion{Fe}{1}  &  7.50  &  4.51  &  $-$2.99  &  0.00  &  0.12  &  27 \\ 
    \ion{Fe}{2}  &  7.50  &  4.51  &  $-$2.99  &  0.00  &  0.12  &  125 \\ 
    \ion{Co}{1}  &  4.99  &  1.98  &  $-$3.02  &  $-$0.02  &  0.10  &  2 \\ 
    \ion{Ni}{1}  &  6.22  &  3.26  &  $-$2.96  &  0.03  &  0.10  &  3 \\ 
    \ion{Cu}{1}  &  4.19  &  $<0.15$  &  $<-$4.04  &  $<-$1.05  &    &  1 \\ 
    \ion{Zn}{1}  &  4.56  &  1.72  &  $-$2.84  &  0.15  &  0.20  &  1 \\ 
    \ion{Ga}{1}  &  3.04  &  $<0.05$  &  $<-$2.99  &  $<0.00$  &    &  1 \\ 
    \ion{Sr}{1}  &  2.87  &  0.88  &  $-$1.99  &  1.00  &  0.20  &  1 \\ 
    \ion{Sr}{2}  &  2.87  &  0.18  &  $-$2.69  &  0.30  &  0.10  &  3 \\ 
    \ion{Y}{2}  &  2.21  &  $-$0.79  &  $-$3.00  &  $-$0.01  &  0.10  &  12 \\ 
    \ion{Zr}{2}  &  2.58  &  $-$0.13  &  $-$2.71  &  0.28  &  0.10  &  4 \\ 
    \ion{Mo}{1}  &  1.88  &  $-$0.95  &  $-$2.83  &  0.16  &  0.20  &  1 \\ 
    \ion{Ru}{1}  &  1.75  &  $-$0.41  &  $-$2.16  &  0.84  &  0.10  &  2 \\ 
    \ion{Ba}{2}  &  2.18  &  $-$0.46  &  $-$2.64  &  0.35  &  0.10  &  3 \\ 
    \ion{La}{2}  &  1.10  &  $-$1.15  &  $-$2.25  &  0.74  &  0.10  &  16 \\ 
    \ion{Ce}{2}  &  1.58  &  $-$0.83  &  $-$2.41  &  0.58  &  0.10  &  13 \\ 
    \ion{Pr}{2}  &  0.72  &  $-$1.39  &  $-$2.11  &  0.88  &  0.10  &  7 \\ 
    \ion{Nd}{2}  &  1.42  &  $-$0.65  &  $-$2.07  &  0.92  &  0.10  &  14 \\ 
    \ion{Sm}{2}  &  0.96  &  $-$0.98  &  $-$1.94  &  1.05  &  0.10  &  11 \\ 
    \ion{Eu}{2}  &  0.52  &  $-$1.19  &  $-$1.71  &  1.28  &  0.10  &  4 \\ 
    \ion{Gd}{2}  &  1.07  &  $-$0.83  &  $-$1.90  &  1.09  &  0.10  &  7 \\ 
    \ion{Tb}{2}  &  0.30  &  $-$1.48  &  $-$1.78  &  1.21  &  0.10  &  4 \\ 
    \ion{Dy}{2}  &  1.10  &  $-$0.66  &  $-$1.76  &  1.23  &  0.10  &  6 \\ 
    \ion{Ho}{2}  &  0.48  &  $-$1.42  &  $-$1.90  &  1.10  &  0.10  &  4 \\ 
    \ion{Er}{2}  &  0.92  &  $-$1.03  &  $-$1.95  &  1.04  &  0.10  &  4 \\ 
    \ion{Tm}{2}  &  0.10  &  $-$1.79  &  $-$1.89  &  1.10  &  0.10  &  4 \\ 
    \ion{Yb}{2}  &  0.84  &  $-$0.80  &  $-$1.64  &  1.35  &  0.20  &  1 \\ 
    \ion{Lu}{2}  &  0.10  &  $-$1.29  &  $-$1.39  &  1.60  &  0.20  &  1 \\ 
    \ion{Hf}{2}  &  0.85  &  $-$1.22  &  $-$2.07  &  0.92  &  0.11  &  2 \\ 
    \ion{Os}{1}  &  1.40  &  $-$0.19  &  $-$1.59  &  1.40  &  0.12  &  3 \\ 
    \ion{Ir}{1}  &  1.38  &  $-$0.31  &  $-$1.69  &  1.30  &  0.20  &  1 \\ 
    \ion{Pb}{1}  &  1.75  &  $-$0.37  &  $-$2.12  &  0.87  &  0.20  &  1 \\ 
    \ion{Th}{2}  &  0.02  &  $-$1.31  &  $-$1.36  &  1.63  &  0.10  &  3 \\ 
    \ion{U}{2}  &  $-$0.54  &  $-$2.13  &  $-$1.59  &  1.40  &  0.20  &  1 \\ 
    \hline
    \end{tabular}
    \end{tightcenter}
    \tablenotetext{a}{Natal $\textrm{[C/Fe]}$=+0.24, based on corrections from \citet{placco2014}.}
    }
    \end{table}

	\begin{figure*}[t!]
	 \begin{center}
	 \begin{minipage}{0.32\textwidth}
	 \includegraphics[width=0.95\textwidth]{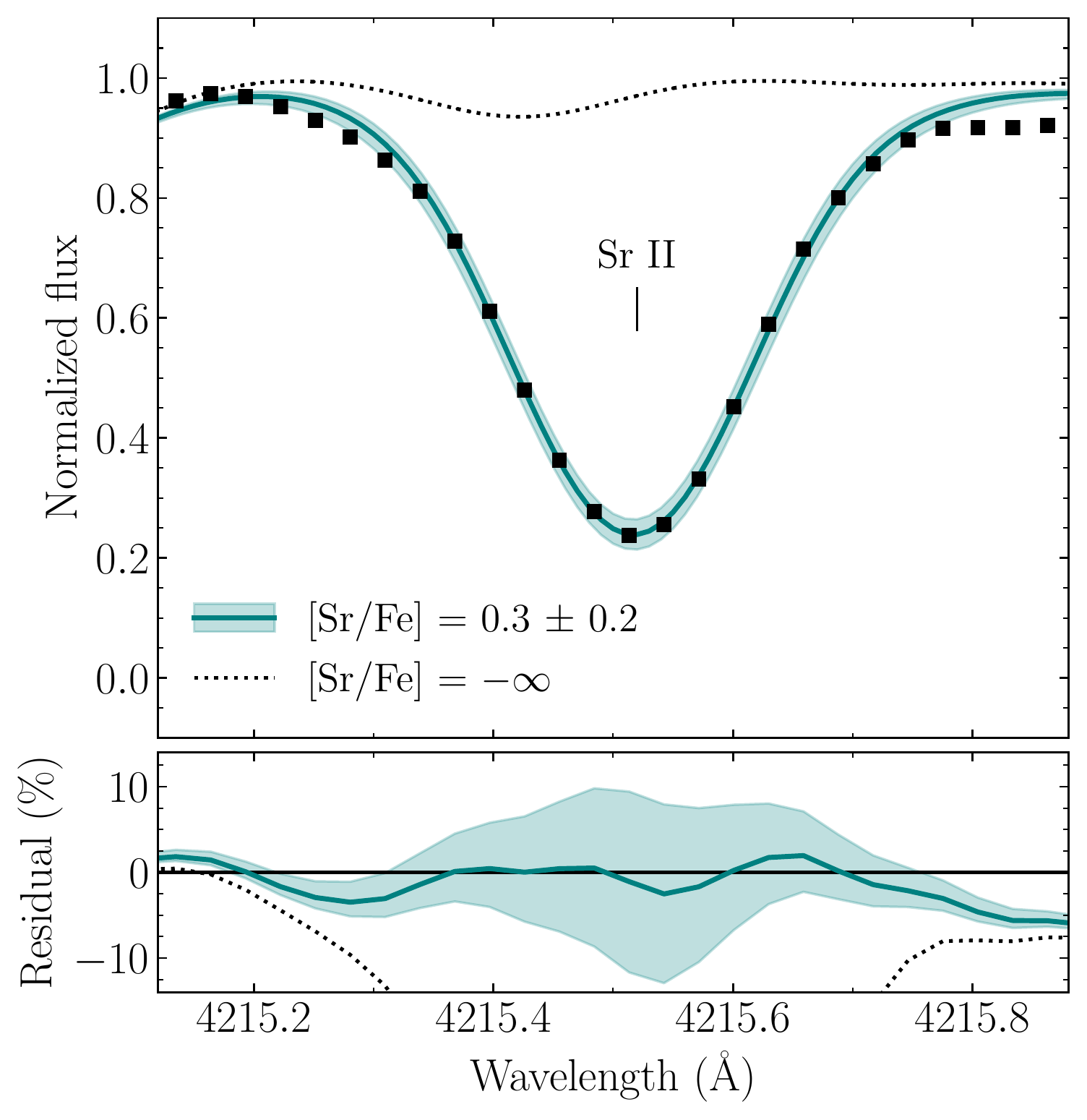}
	\end{minipage}
	 \begin{minipage}{0.32\textwidth}
	 \includegraphics[width=0.95\textwidth]{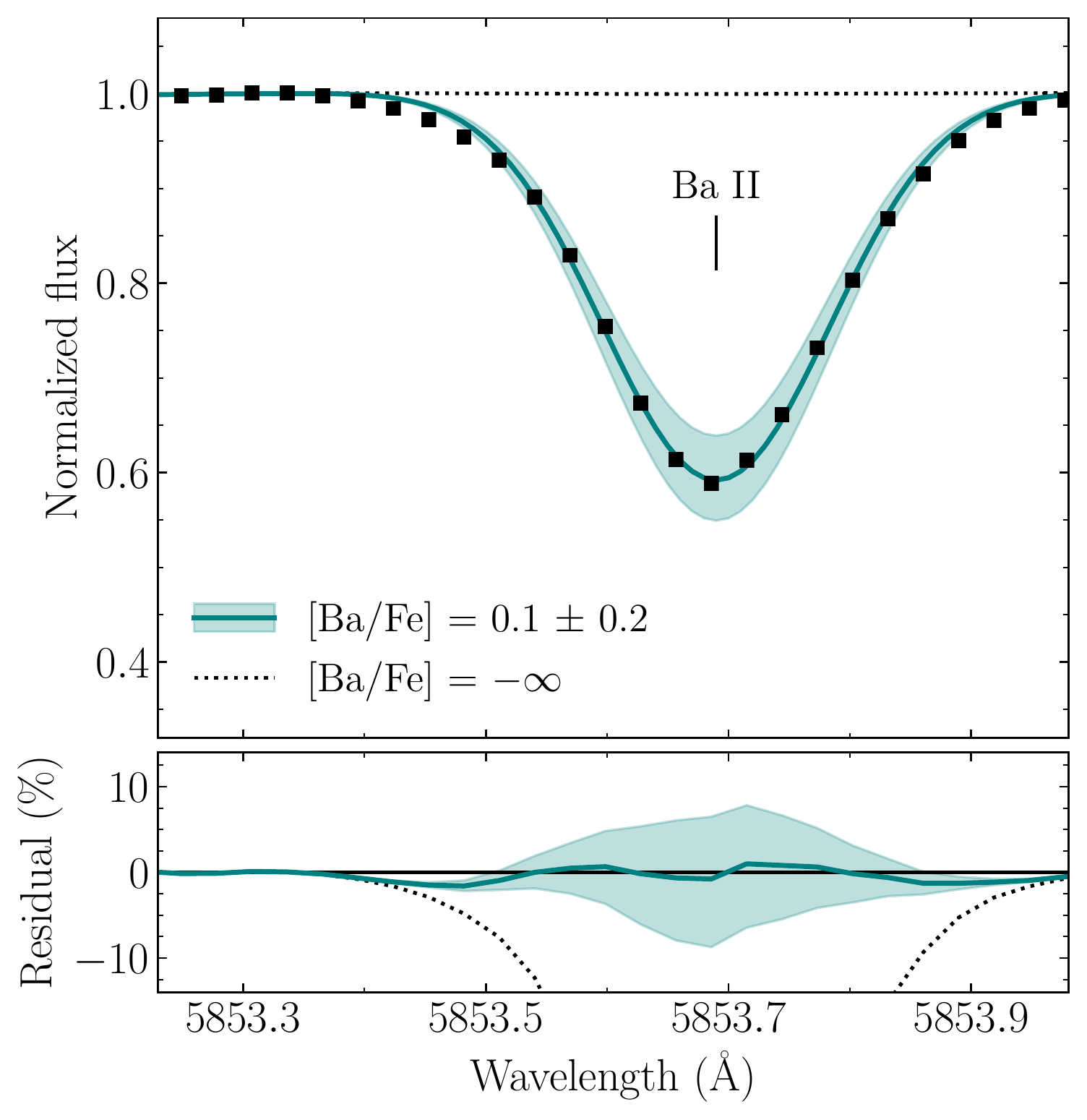}
	\end{minipage}
	\begin{minipage}{0.32\textwidth}
	 \includegraphics[width=0.95\textwidth]{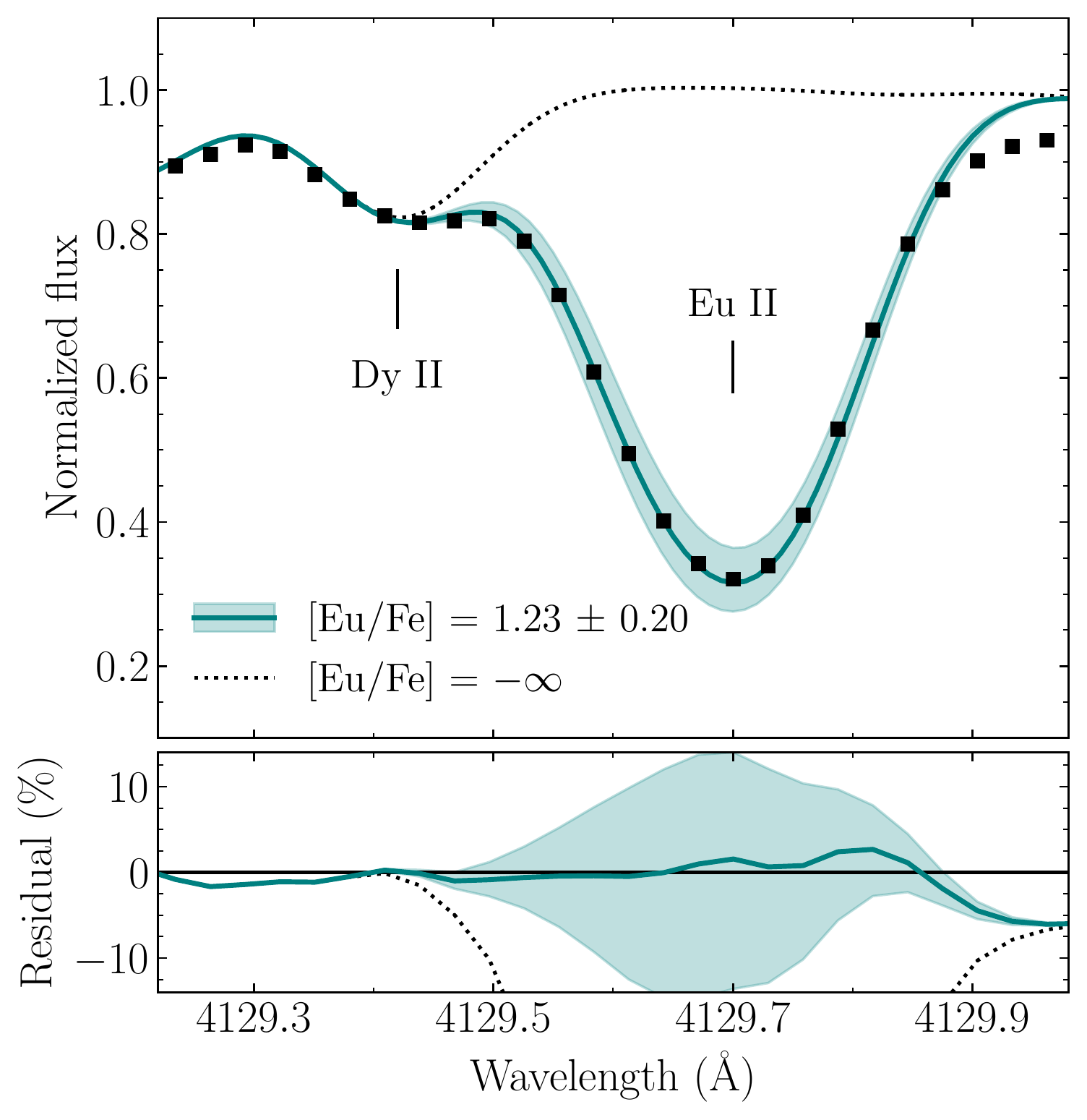}
	\end{minipage}\\ \vspace{2mm} \hfill
	 \begin{minipage}{0.32\textwidth}
	 \includegraphics[width=0.95\textwidth]{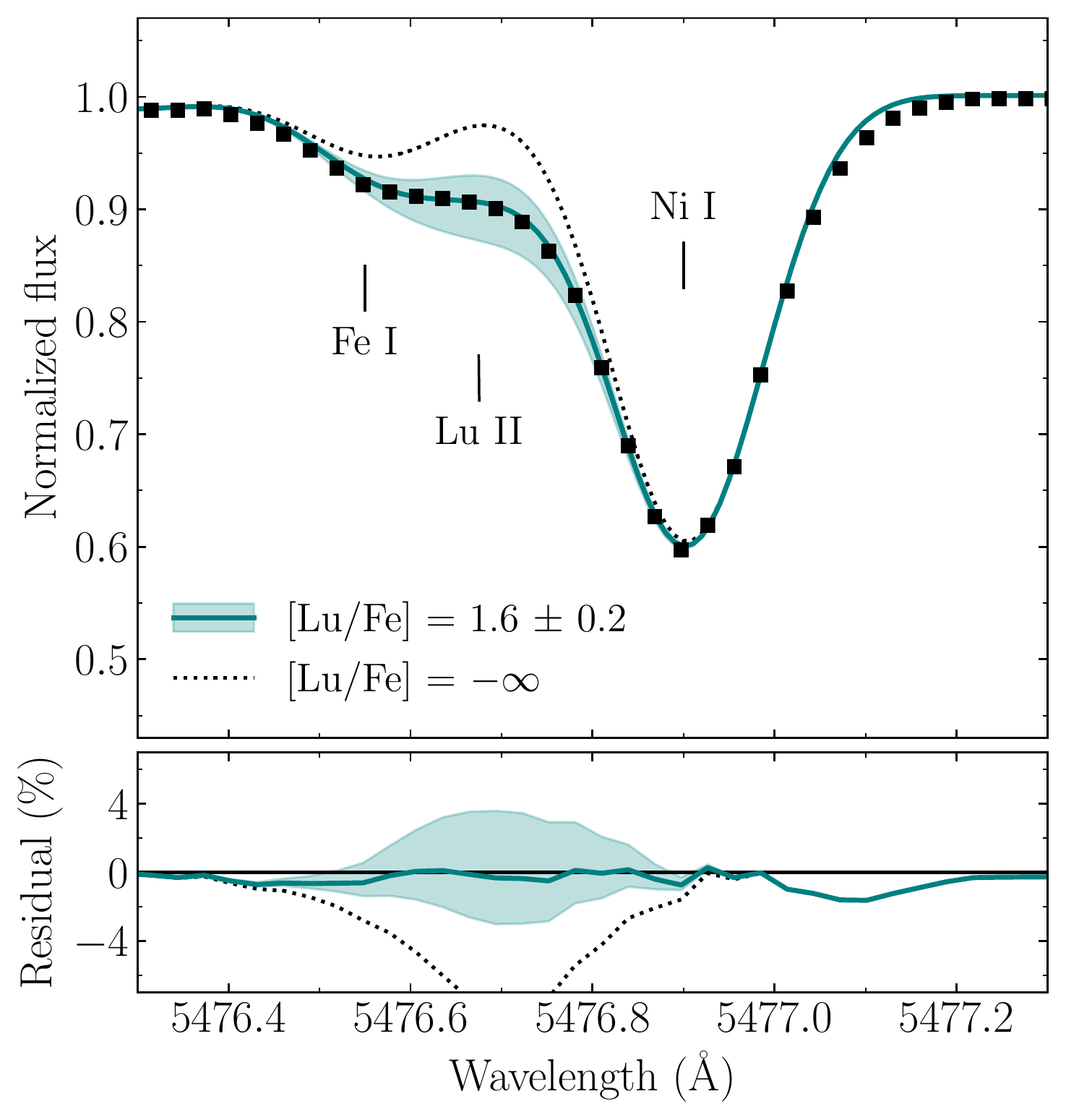}
	\end{minipage}
	 \begin{minipage}{0.32\textwidth}
	 \includegraphics[width=0.95\textwidth]{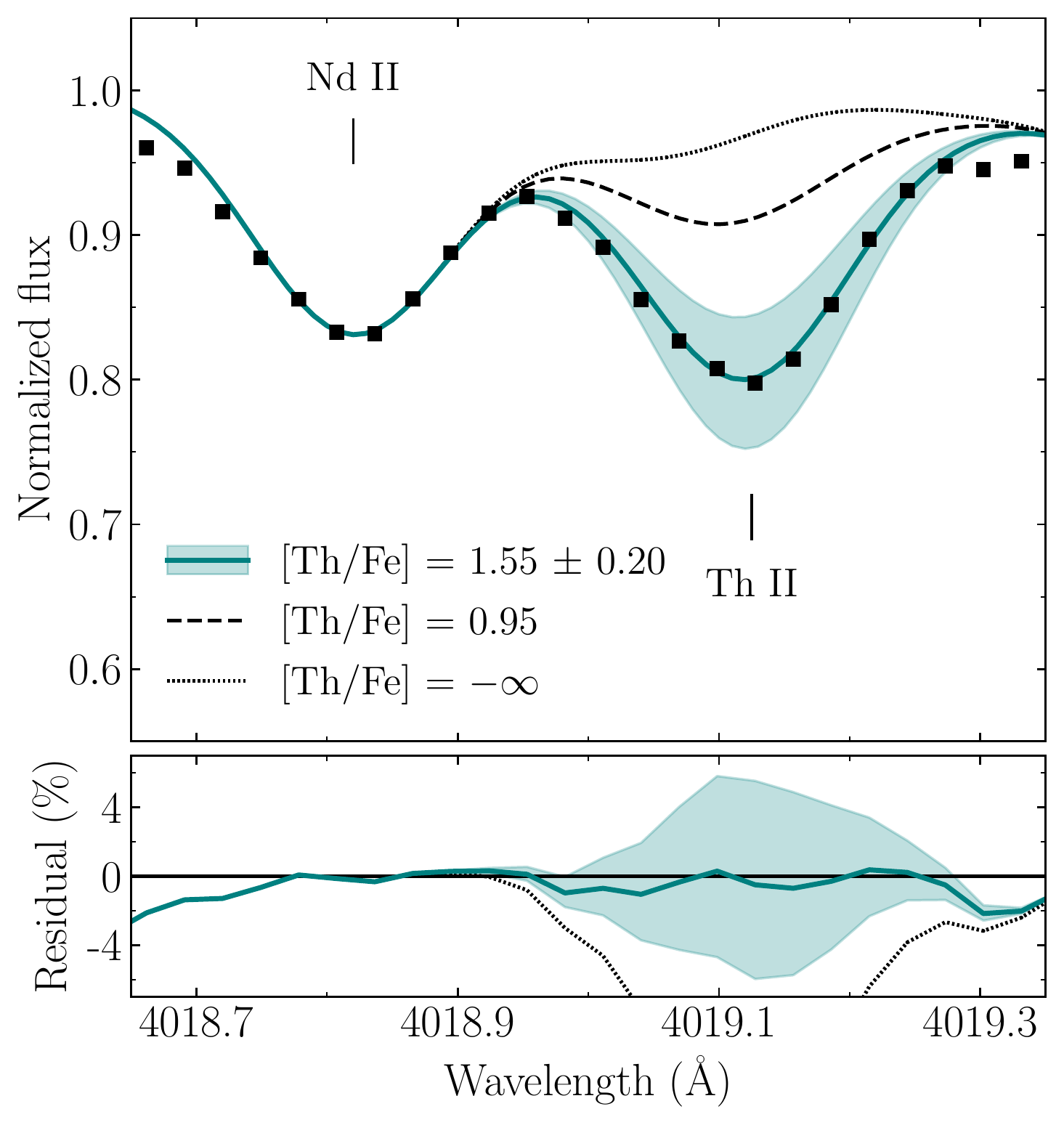}
	\end{minipage}
	\begin{minipage}{0.32\textwidth}
	 \includegraphics[width=0.95\textwidth]{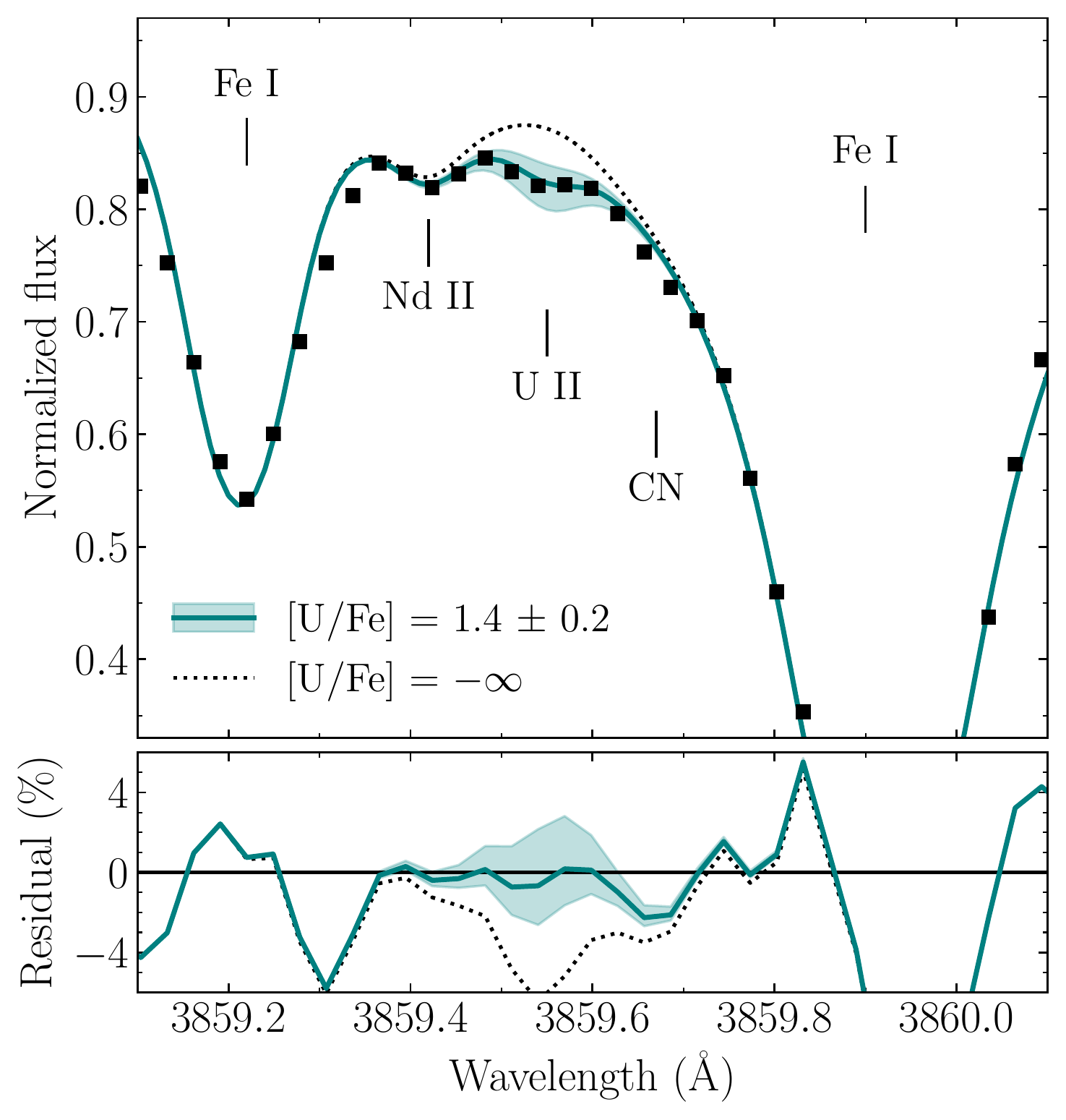}
	\end{minipage}
	 \caption{\label{fig:syntheses} Relevant syntheses and derived abundances for important \rp\ elements present in \shortname. The top panels of each plot show the best-fit syntheses (solid lines) and 0.2~dex uncertainties (shaded regions) compared to the observed spectrum (points). Key features are labeled. The bottom panels show the residuals between the observed spectrum and the synthetic fits. Derived abundances corresponding to the plotted syntheses are indicated in each legend. The $\textrm{[Th/Fe]}=0.95$ synthesis corresponds to a typical non-actinide-enhanced \ion{Th}{2} line profile.}
	 \end{center}
	\end{figure*}

Derived neutron-capture-element abundances of \shortname\ are shown in Figure~\ref{fig:pattern}.
The abundances unequivocally match a scaled-Solar pure \rp\ pattern, with no apparent contribution from the \emph{s}-process.
Also shown in Figure~\ref{fig:pattern} is a comparison to the elemental-abundance pattern of the canonical actinide-boost star, CS~31082--001 \citep{hill2002}.
Table \ref{tab:abundances} lists the derived abundances for all elements identified in \shortname, their comparison to \citet{asplund2009} Solar abundances, the uncertainty of the abundance, and the number of lines measured.
A complete linelist of all measured transitions and derived individual abundances is provided in the .tar.gz package.
We adopt a minimum uncertainty of 0.10~dex.
For elements with only one feature, a 0.2~dex uncertainty is assumed.
Important spectral features, including Sr, Ba, and Eu---used to classify \rp-enhancement in metal-poor stars---are shown in Figure~\ref{fig:syntheses}.
Abundances of the elements from C to Ga in \shortname\ resemble those of normal metal-poor halo stars.
In the following, we focus on the neutron-capture element abundances.

As with all \rii\ stars, the heavy-element abundance pattern of \shortname\ reproduces the observed scaled-Solar lanthanide abundances, with deviations greater than the formal uncertainties for only a few elements.
However, the most striking difference between \shortname\ and other \rii\ stars is its actinide signature.

\subsection{Light Neutron-Capture Elements}

The elements Sr, Y, Zr, Mo, and Ru were measured in this star; atomic transitions of Rh, Pd, and Ag occur at wavelengths bluer than the instrumental system limit for efficient throughput, preventing measurement of their abundances.
It has been previously noted that a significant spread in the light neutron-capture elements exists in \rp-enhanced stars, with \ri\ stars generally exhibiting stronger relative first-peak enhancement \citep{siqueira2014}.
This spread is often attributed to a separate \rp, which is referred to as a limited, or weak, \rp\ \citep{wanajo2006,hansen2012,frebel2018}, and is distinct from that responsible for the production of second- and third-peak elements.
Compared to the scaled-Solar pattern, the light \rp\ elements in \shortname\ are indeed slightly lower, indicating little limited-\emph{r} contribution.
It should be noted that Mo has a significant \emph{p}-process component, which is not accounted for in the Solar \rp\ residual, likely causing its apparent under-abundance \citep{meyer1994}.

\subsection{Heavy Neutron-Capture Elements}
	
We were able to derive abundances for all stable elements between Ba and Hf, as well as Os, Ir, and Pb.
Like most \rii\ stars, the abundance pattern of \shortname\ in this region agrees well with scaled-Solar \rp\ values, with some deviations.
Intriguingly, the abundance of \ion{Yb}{2} is much higher than the scaled-Solar \rp\ value in \shortname; this over-abundance is also observed in CS~29497--004 \citep{hill2017} and CS~22892--052 \citep{sneden2008}, neither of which display an actinide boost.
However, it should be noted that the Yb abundance is derived from only one feature, and is affected by hyperfine splitting, which is included in this synthesis.
Similarly, Lu and Ir appear over-abundant relative to the Solar \rp\ pattern, but these elements each have just one optical feature suitable for abundance derivations.
Figure~\ref{fig:syntheses} shows the synthesis of the lutetium feature from which we derived the \ion{Lu}{2} abundance.

\subsection{Thorium, Uranium, and the Actinide Boost}

	\begin{figure}[t]
	 \begin{center}
\centerline{\includegraphics[width=\columnwidth]{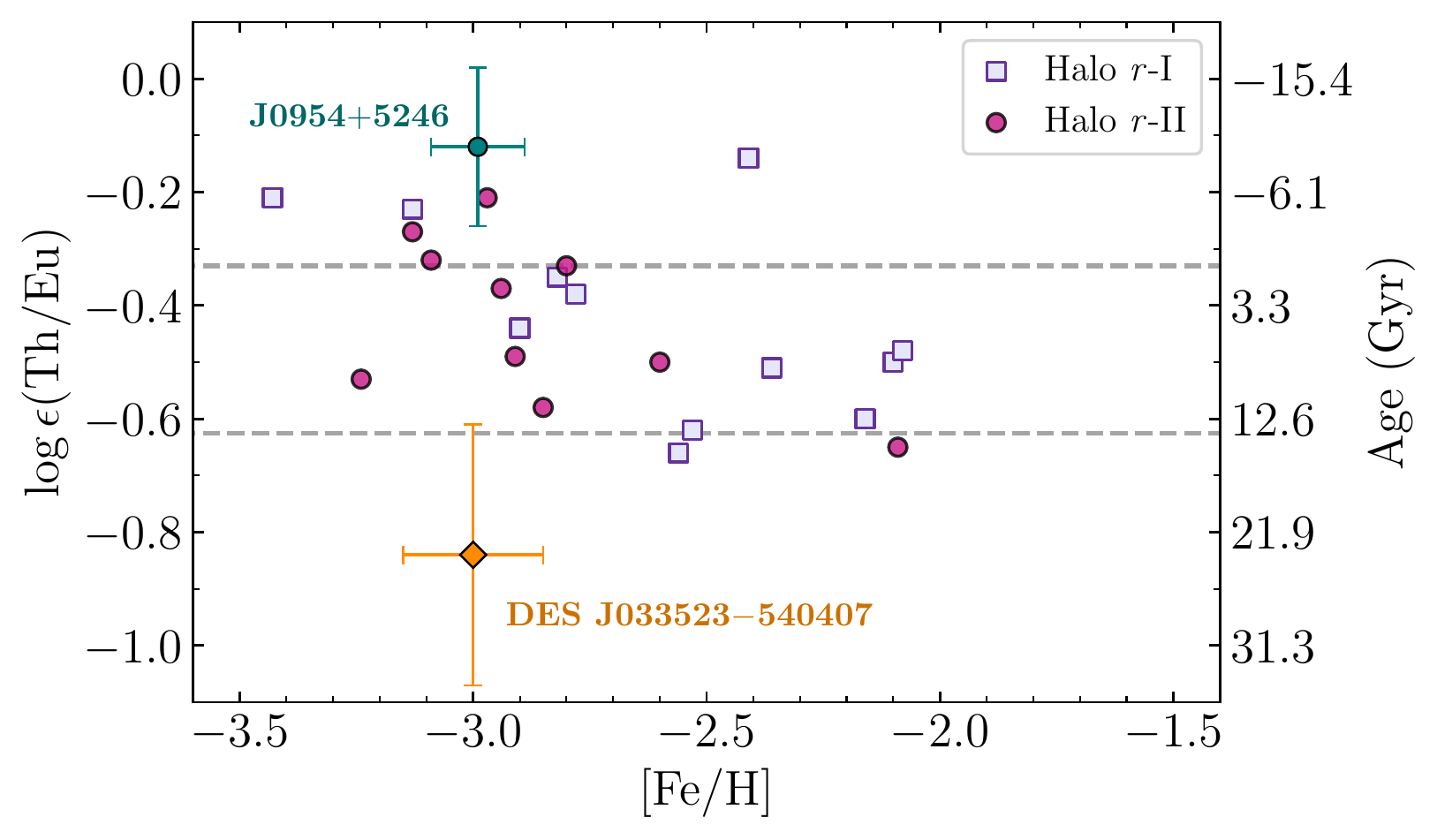}}
	 \caption{\label{fig:thstars} Th/Eu abundances vs. metallicity for known \ri\ and \rii\ stars where a Th measurement is provided. Also shown is the Reticulum II member, DES~J033523$-$540407, reported by \citet{ji2018}. The right axis shows the Th/Eu age calculated based on \citet{schatz2002} production ratios. Horizontal lines indicate physical limits on the age (i.e., 0 and 13.8 Gyr). Data were generated using JINAbase \citep{abohalima2017}, supplemented by measurements from \citet{mashonkina2014}, \citet{roederer2014}, \citet{hill2017}, \citet{placco2017}, and \citet{sakari2018}.}
	 \end{center}
	\end{figure}

Uranium is among the most difficult elements to measure in a star.
However, being able to measure uranium---together with thorium---provides clues to the nature and possible site of the \rp\ that synthesized the actinides.
Three lines of \ion{Th}{2} were measured in this star, spanning an abundance range of 0.23~dex.
Only one line of \ion{U}{2} was measured; syntheses are shown in Figure~\ref{fig:syntheses}.
The actinide-to-lanthanide ratio of $\text{[Th/Eu]} = +0.37$ demonstrates the actinide-boost designation of \shortname.

Figure~\ref{fig:thstars} shows the $\log\epsilon(\textrm{Th/Eu})$ abundances of halo \ri\ and \rii\ stars; non-actinide-boost stars typically have $\log\epsilon(\textrm{Th/Eu})$ values near $-0.5$.
Even allowing for a 0.15~dex uncertainty, the $\log\epsilon(\textrm{Th/Eu})$ abundance of \shortname\ is larger than its non-actinide-boost cousins by a factor of $\sim$2, making it the most actinide-enhanced \rii\ star currently known.
The actinide enhancement of \shortname\ is even larger than that of the canonical actinide-boost star CS~31082$-$001 \citep[\text{[Th/Eu]} = +0.20;][]{hill2002}.

\section{Discussion and Conclusion}

The scatter of the Sr-Y-Zr-group abundances among \rp-enhanced stars may be attributed to different levels of production by the limited \rp, and thus the production of the first \rp\ peak can be decoupled from the second and third peaks.
However, it seems unlikely that the actinides can be similarly decoupled from the main \rp, or attributed to a separate \rp\ entirely.
The presence of the actinide boost demonstrates that a real variation exists among the actinide abundances.
Therefore, measuring Th (and U) abundances for a larger number of \rp-enhanced stars is essential to distinguish between possible astrophysical \rp\ sites and to determine what conditions enable strong actinide production.
Specifically, actinide enhancement could indicate fission cycling in \rp\ nucleosynthesis, which may only occur under certain astrophysical conditions.
We are currently exploring the effect of fission cycling in the low-entropy dynamical ejecta of a neutron star merger as one possibility for the origin of the actinide-boost phenomenon (Holmbeck et al., in prep.).

On the other hand, there may exist evidence for \emph{weak} actinide production.
\citet{ji2018} measured thorium for the brightest \rii\ star in the UFD galaxy Reticulum II: DES~J033523$-$540407.
This star exhibits an \emph{under}-abundance of thorium, relative to europium, even though the rest of the main \rp\ is consistent with and other \rp-enhanced stars.
The $\log\epsilon(\textrm{Th/Eu})$ abundance of DES~J033523$-$540407 is shown relative to its Milky Way counterparts in Figure~\ref{fig:thstars}.
Such a low thorium abundance has not yet been observed in Milky Way halo stars, which brings into question the type of \rp\ that could under-produce the actinides.

The presence of long-lived radioactive actinides allows an approximate age determination by radioactive decay dating, as described in \citet{placco2017}.
In the absence of a uranium measurement, the Th/Eu chronometer is used to derive ages by comparing to a set of initial production ratios from \rp\ simulations, e.g., from \citet{schatz2002}.
However, in actinide-boost stars, using the Th/Eu ratio leads to unphysical values, since the measured Th/Eu ratio may be higher than the theoretical initial production ratio.
To our knowledge, none of the existing \rp\ nucleosynthesis models can account for the abundance pattern observed in actinide-boost stars.
Notwithstanding this shortcoming, the Th/U chronometer can still be used to estimate stellar ages since the actinide boost affects thorium and uranium proportionally.
The Th/U chronometer is insensitive to the boost and thus leads to more realistic ages of actinide-boost stars than the Th/Eu chronometer.

Using the production ratios from \citet{schatz2002}, the Th/U age is 13.0$\pm$4.7~Gyr, commensurate with its low metallicity.
Uncertainties on the age contain only the measured abundance uncertainty and do not include those on the production ratios.
The Th/Eu and U/Eu ages are $-$9.5~Gyr and 5.8~Gyr, respectively.
These ages can be made to agree with the Th/U age of 13.0~Gyr by boosting the initial production ratios of thorium and uranium by a factor of 3.1.
In a similar analysis, \citet{schatz2002} found a ``boost factor" of 2.5 for CS~31082--001.

Values for this boost factor vary, as Th/Eu abundances are not found at discrete enhancement levels, but rather, cover a range.
With refined stellar main \rp\ abundances---particularly for Th, U, and the third \rp-peak elements, where few features exist from which to derive abundances---as well as species that can only be measured from UV data, we may find clues that are unique to and correlated with actinide production, and thereby characterize the actinide boost and better understand its origin.

\acknowledgments
E.M.H., T.C.B., I.U.R., and V.M.P. acknowledge partial support for this work from grant PHY 14-30152; Physics Frontier Center/JINA Center for the Evolution of the Elements (JINA-CEE), awarded by the US National Science Foundation.
This work has been supported in part by NSF grant AST1616040 to C.S.
Y.S.L. acknowledges support from the National Research Foundation of Korea grant funded by the Ministry of Science and ICT (No.2017R1A5A1070354, NRF-2015R1C1A1A02036658, and NRF-2018R1A2B6003961).

\facilities{Smith, LAMOST}

\software{\texttt{IRAF} \citep{tody1993}, \texttt{linemake} (\url{https://github.com/vmplacco/linemake}), \texttt{MOOG} \citep[\url{https://github.com/alexji/moog17scat};][]{sneden1973,sobeck2011}, \texttt{Matplotlib} \citep{hunter2007}}

\bibliographystyle{apj}

\end{document}